\documentclass[showpacs,prb,twocolumn]{revtex4}
\usepackage{amsmath}
\usepackage{amssymb}
\usepackage{color}
\usepackage{graphics}
\usepackage{graphicx}
\usepackage{dcolumn}

\begin{document}

\title {Edge and waveguide THz surface plasmon modes in graphene micro-ribbons}

\author{ A.~Yu.~Nikitin$^{1,2}$}
\email{alexeynik@rambler.ru}
\author{ F. Guinea$^3$}
\author{ F.~J.~Garc\'{i}a-Vidal$^4$}
\author{ L.~Mart\'{i}n-Moreno$^1$}
\email{lmm@unizar.es}
 \affiliation{$^1$ Instituto de Ciencia de Materiales de Arag\'{o}n and Departamento de F\'{i}sica de la Materia Condensada,
CSIC-Universidad de Zaragoza, E-50009, Zaragoza, Spain \\
$^2$A.Ya. Usikov Institute for Radiophysics and Electronics, Ukrainian Academy of
Sciences, 12 Acad. Proskura Str., 61085 Kharkov, Ukraine\\
$^3$ Instituto de Ciencia de Materiales de Madrid, CSIC, Cantoblanco,
E-28049 Madrid, Spain\\
$^4$ Departamento de F\'{i}sica Te\'{o}rica de la Materia Condensada, Universidad Aut\'{o}noma de Madrid, E-28049
Madrid, Spain}

\begin{abstract}

Surface plasmon modes supported by graphene ribbon waveguides are studied and classified.
The properties of both modes with the field concentration within the ribbon area (waveguiding modes) and on the edges (edge modes) are discussed. The waveguide and edge modes are shown to be separated from each other by a gap in wavenumbers. The even-parity hybridized edge mode results to be the fundamental electromagnetic mode of the ribbon, possessing also the lowest losses. All the plasmonic modes in the ribbons have an optimum frequency, at which the absorption losses are minimum, due to competition between the plasmon confinement and the frequency dependence of absorption in graphene.

\end{abstract}

 \maketitle

The science of graphene is advancing rapidly, discovering new fundamental physical effects\cite{Review09} and showing promising perspectives in several state-of-the-art technological applications\cite{graphphot_natphot10}. For example, graphene can carry both electromagnetic (EM) signals and electric currents through the same extremely thin circuitry, which could allow electrical interconnects to achieve data transmission faster rates. In this regard, it has been recently shown that graphene ribbons can operate as a broadband radio-frequency mixer at frequencies up to 10 GHz inside an integrated circuit\cite{THzcircuit}, showing a great potential for waveguiding. The two main characteristics of graphene for this kind of applications are: (i) that despite being almost transparent, graphene can support surface plasmons (GSPs) in the THz regime\cite{IRSPP09,THzSPP11} and (ii) the carrier concentration (and thus the conductivity) in graphene can be controlled through electrostatic gating. This last property can, in turn, be used to tune the properties of GSPs, opening up a wealth of interesting applications in photonics\cite{Hansonw08,ribbonsPRB10,graphphot_natphot10,Bludov10,p-n_junction,Engheta,deAbajo2011,Nikitin2011}.

For most functionalities the GSPs must ideally also be confined laterally in the graphene sheet, which is known to occur in graphene ribbons\cite{BreyPRB07,SilvestrovPRB08,ribbonsPRB10,p-n_junction,Engheta}. However, a characterization of the confined plasmons in this geometry is lacking. In this paper we address this problem, and study the characteristics of GSP modes in graphene ribbons of micrometric widths, in the THz regime. We concentrate in the spectral regimes where the ribbon supports either a single EM mode or a few of them and, in this last case, on the properties of strongly localized edge GSP modes (EGSP) that appear in the system.  The dependence of both waveguide GSP (WGSP) and EGSP upon relaxation time of electrons, permittivity of a substrate and width of the ribbon, is also analyzed.

\begin{figure}[thb!]
\includegraphics[width=6cm]{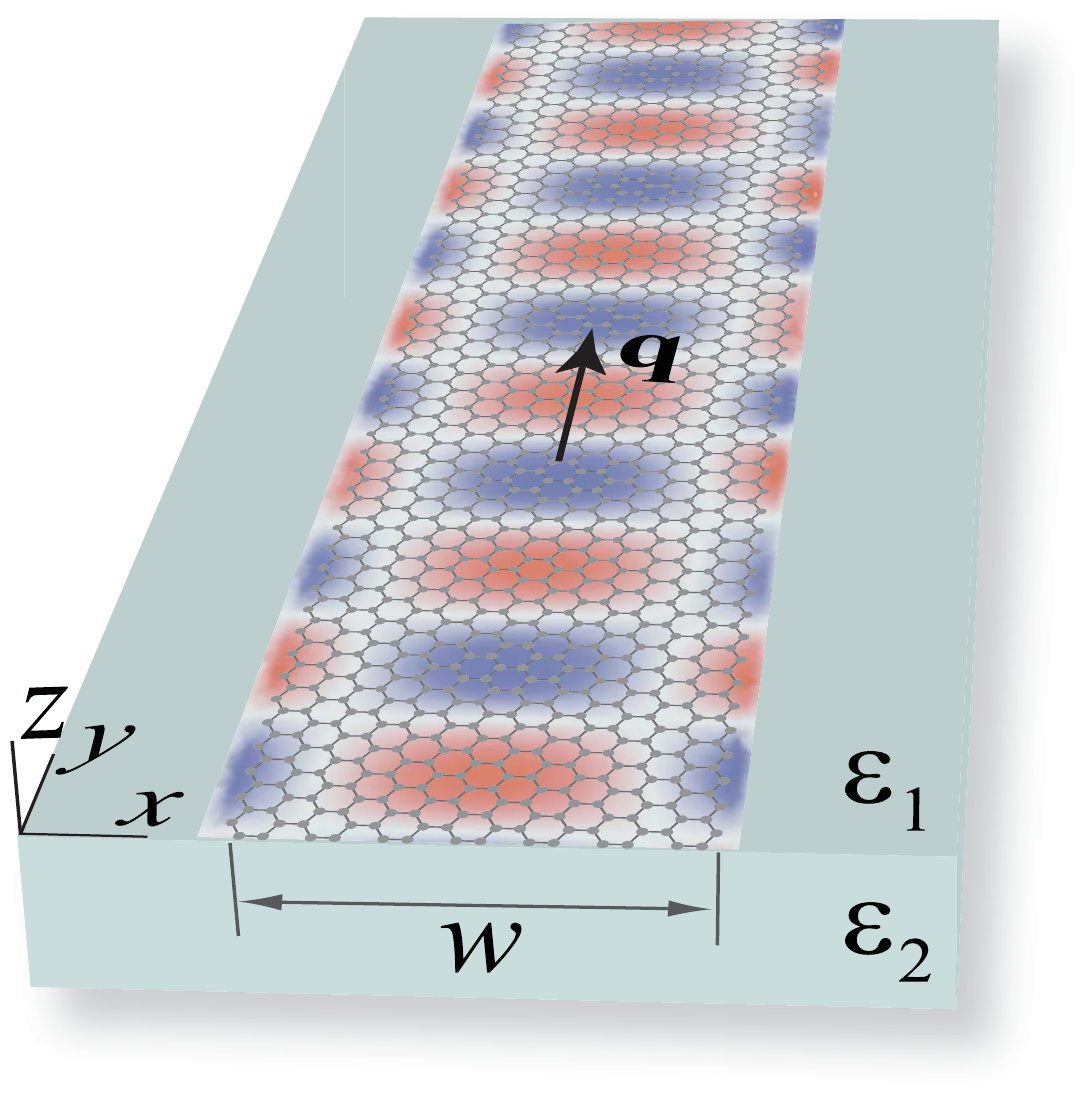}\\
\caption{(Color online)  The geometry of the studied system: a graphene ribbon of the width $w$, and conductivity $\sigma$, placed at the interface between two dielectric half-spaces with dielectric constants $\varepsilon_1$ and $\varepsilon_2$. The mode propagates along the $Oy$ axis.}\label{geom}
\end{figure}

Let us consider a graphene ribbon of width $w$ (at $|x|<w/2$), placed at the boundary $z=0$ between two dielectric half-spaces, see Fig.~\ref{geom}.  The (frequency-dependent) two-dimensional conductivity of graphene is $\sigma(\omega)$. The ribbon can be either an actual graphene strip, or  ``virtually'' created by spatially varying external gates acting on a graphene sheet, as proposed in Ref.~\onlinecite{Engheta}. Here we only report on  the case where $\sigma=0$ outside the ribbon, but we have checked that our results are virtually unmodified if this constrain is relaxed, provided GSPs are not supported by graphene for $|x|>w/2$.

\begin{figure}[tbh!]
  \includegraphics[width=8.5cm]{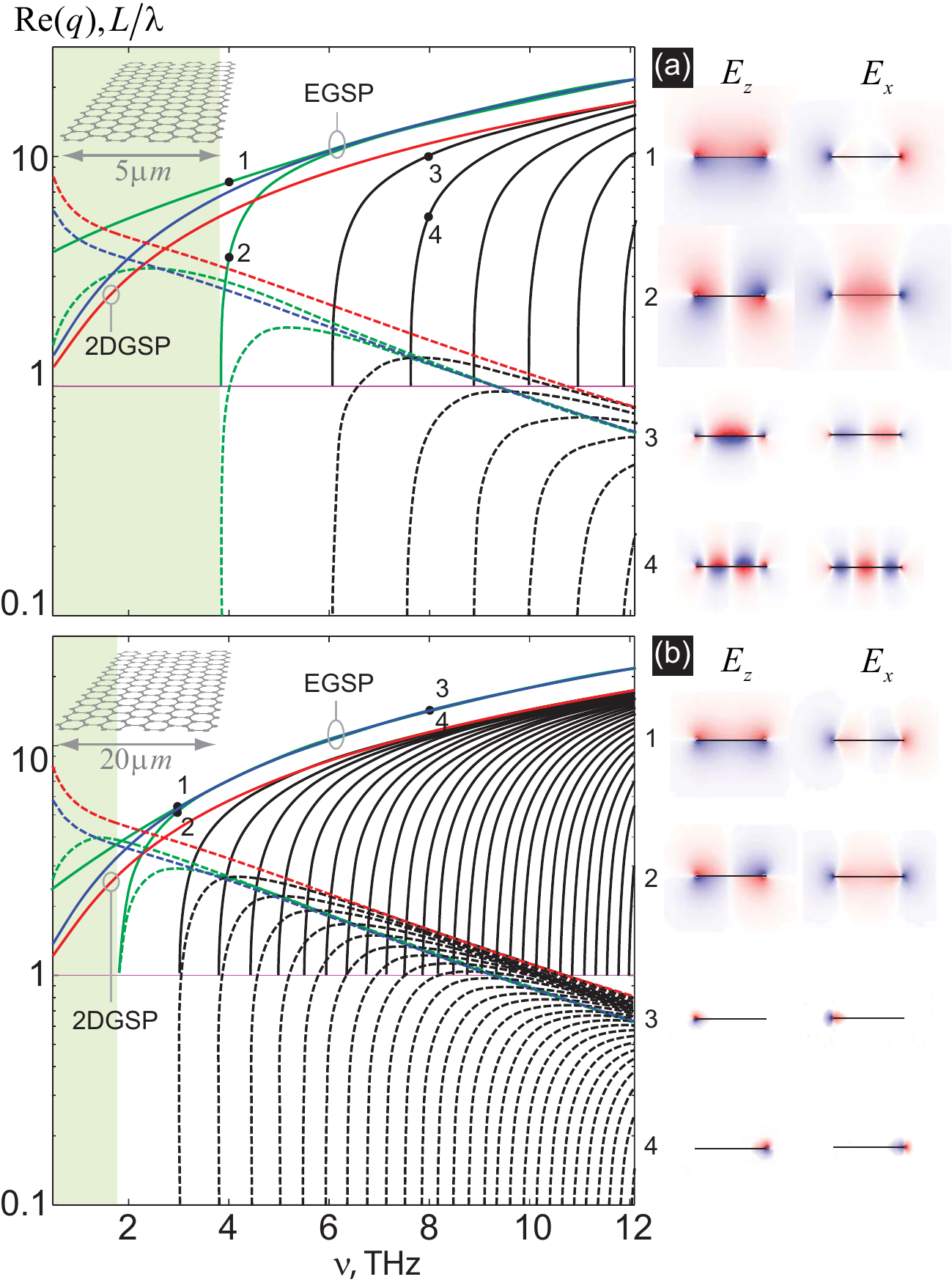}\\
\caption{(Color online) Normalized wavevector of GSP modes (continuous curves) and propagation lengths (discontinuous curves) as a function of frequency for free-standing graphene ribbons of 5 $\mu$m (a) and 20 $\mu$m (b) widths. The curve  for dispersion law  ``EGSP'' is for edge GSPs in a semi-infinite graphene sheet, while the one marked as ``2DGSP'' is for GSP in an infinite graphene sheet. The colorplots for $x$- and $z$-components of the electric field are shown in the vicinity  of the ribbon cross-section. The numbers next to the colorplots correspond to the labels on the dispersion curves.}\label{disp}
\end{figure}

Due to translational symmetry, the electric field of each EM eigenmode in the ribbon has the form
$\vec{E}(\vec{r},t) = \vec{E}(x, z) \, \mathrm{exp}(i q k_\omega y) \, \mathrm{exp}(-i \omega t)$, where $k_\omega=\omega/ c$ is the free space wavevector, $c$ is the speed of light and $q(\omega)$ is the modal wavevector in dimensionless units. We will refer to the real part of $ q(\omega)$ as the "normalized wavevector", while its imaginary part provides the propagation length of GSPs, $L$, thorough $L/\lambda = 2 \pi / \mathrm{Im}(q)$, where $\lambda = 2 \pi / k_\omega$ is the wavelength in vacuum.

For the calculations, we use the conductivity computed within the standard random phase approximation \cite{Wunsch06,Hwang07,Falkovsky08}, which depends on temperature ($T$), chemical potential ($\mu$) and scattering time ($\tau$). We chose $T=300$K, $\mu=0.2$ eV and, unless otherwise stated, a relaxation energy $E_\tau = 2 \pi \hbar/ \tau = 0.1meV$ (corresponding to a mobility of $1.87 \times 10^6 \, cm^2 V^{-1} s^{-1}$). All calculations have been performed by using the finite-elements commercial software ``COMSOL''.

\begin{figure}[tbh!]
  \includegraphics[width=6cm]{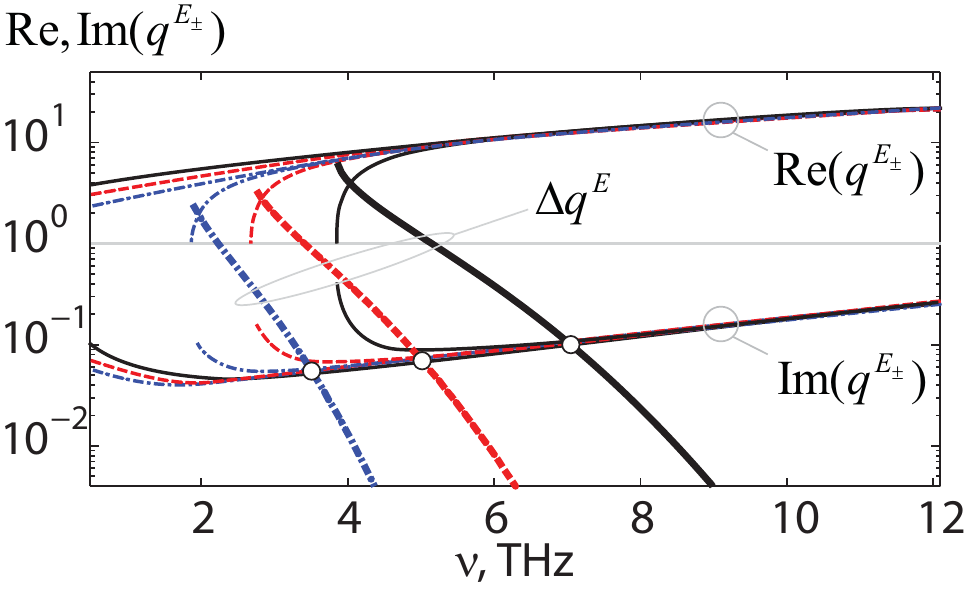}\\
\caption{(Color online) Difference between the normalized wavevectors of the two coupled EGSP ribbon modes, $\Delta q$, together with $\mathrm{Re}(q^{E_\pm})$ and
$\mathrm{Im}(q^{E_\pm})$  (corresponding to the edge EM modes of a semiinfinite graphene sheet). The continuous, dashed and dash-dotted curves correspond to $w = 5, 10$ and 20 $\mu$m, respectively. The white circles indicate the points where $\mathrm{Im}(q^{E}) = \Delta q$, when the modes can be considered as degenerated.}\label{split}
\end{figure}

In Fig.~\ref{disp} both the dispersion relations and the propagation lengths of GSPs are rendered, for two widths of free-standing graphene ribbons. These widths have been chosen so that, in the considered frequency range, $w/\lambda<1$ for the  5 $\mu$m-width waveguide and $w/\lambda\lesssim1$ for the 20 $\mu$m-width one. Notice that the dispersion relation is given in the form $q(\nu)$ with $\nu=2\pi\omega$, so that $q=1$ sets the position of the light cone. The normalized wavevectors are shown by continuous curves, while the propagation lengths are represented by discontinuous ones.

We would like to notice that the microscopic structure of graphene edges modifies significantly the electronic spectrum, especially at energies near the Dirac point. The calculations reported here are not affected by the microscopic details of the edges. We consider graphene stripes where the separation between the Fermi energy and the Dirac point is much larger than other scales, like temperature or relaxation energy. At these energies, the main differences between different types of edges is the amount of intervalley scattering that they induce, which vanishes for a zigzag edge and is maximum for armchair edges. The macroscopic analysis discussed here can be written in terms of contributions from the two valleys. However, plasmons are collective excitations of the total electronic charge. An electric current from a given valley incident at a zigzag edge is reflected in the same valley, while an armchair edge leads to a change of valley upon reflection. The total current is, obviously, conserved in both cases. As the calculation depends only on total charges and total currents, the nature of the edge cannot change the results. An influence of the edge structure can be expected, however, near the Dirac point, where the combination of quantum confinement and microscopic structure leads to changes in the electronic spectrum.

The dispersion relation for GSP modes in ribbons can be related to that for (i) 2DGSP modes in an infinite graphene sheet (with wavevector $q^{2D}=\sqrt{1-1/\alpha^2}$, where $\alpha = 2 \pi \sigma/c$), see curve ``2DGSP'' in Fig.~\ref{disp} and (ii) surface plasmon modes appearing at the edges in a semi-infinite graphene sheet\cite{Engheta} (EGSPs) with wavevector $q^{E}(\omega)$ (curve ``EGSP'' in Fig.2). Notice that the dispersion curve for a EGSP lies above that for a 2DGSP, which implies that the former is more tightly confined.

We find that in a ribbon there are two modes that originate from the hybridization of EGSPs. Due to splitting, one of them (the one corresponding to even parity of $E_z$ with respect to the ribbon axis), has a wavevector larger than $q^{E}$ and is the fundamental one in the ribbon, while the other one has $q<q^{E}$.  As seen from this figure, for both cases there is a single-mode region (shaded) for low frequencies. The wavevector of the fundamental mode is larger than that of 2DGSPs, and its propagation length is smaller, so $L$ remains of the order of 10 GSP wavelengths throughout the considered frequency range. For sufficiently large frequencies ($\nu=3.8$ THz for $w=5\mu$m and $\nu=1.8$ THz for $w=20 \mu$m) a second GSP mode is sustained by the ribbon. As the frequency increases, the dispersion relations for these two GSP modes approach each other, merging into the dispersion relation for an EGSP of a semi-infinite sheet. For a given frequency, the smaller the width of the ribbon, the more the edge modes overlap and, correspondingly, the larger the splitting between their wavevectors $q^{E+}$ and $q^{E-}$. The highest frequency for which the two edge modes can be considered as coupled can be estimated from the relation $\Delta q^E< \mathrm{max}\left\{\mathrm{Im}(q^{E_\pm})\right\}\simeq \mathrm{Im}(q^{E})$, which means that the ``line-width'' (due to absorption) of the isolated EM edge mode prevails the splitting (see Fig.~\ref{split}).

\begin{figure}[tbh!]
  \includegraphics[width=7cm]{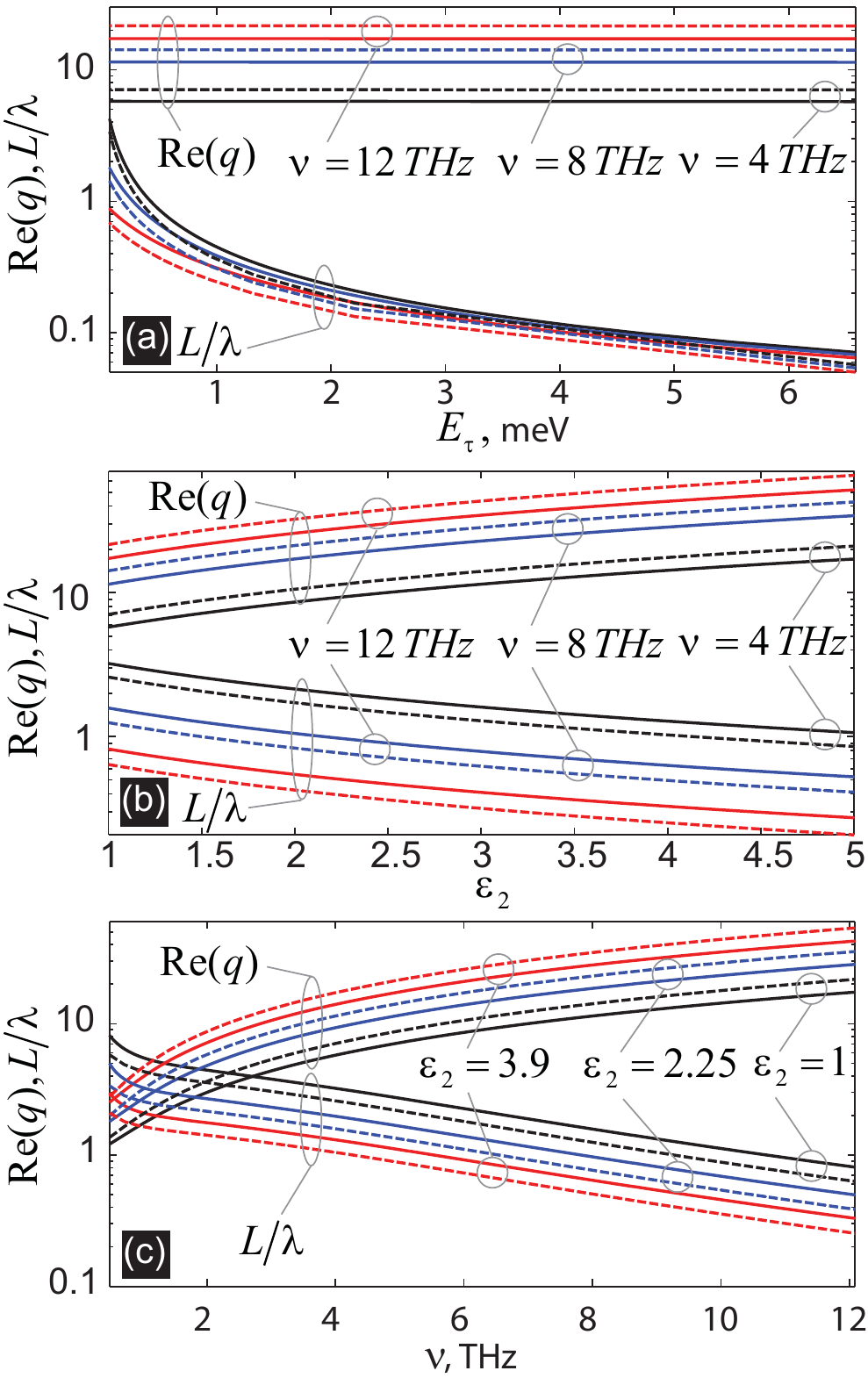}\\
\caption{(Color online) Dependencies of $\mathrm{Re}(q)$ and $L$  (discontinuous lines: EGSP of a semi-infinite sheet, continuous lines:  2DGSP) with:  (a) scattering time (in units of meV), (b) dielectric constant of the substrate, (c) frequency. Unless otherwise stated, $\varepsilon_2=1$ and $E_\tau=0.1$meV.}\label{disp1}
\end{figure}

Additionally to these hybridized edge modes, the ribbon may support several WGSPs having a field that extends over the ribbon width. The number of WGSPs trapped by the ribbon can be estimated as $N_w \sim 2 q^{2D} w/\lambda$, which counts basically how many GSP half-wavelengths fit across the ribbon. These "bulk" modes present a dependence for $q(\omega)$ that lies below the corresponding one for a 2DGSP, which is their high frequency asymptote, see Fig.~\ref{disp}.  Nevertheless, the propagation length of these bulk modes is always smaller that the one for the 2DGSPs (see Fig. ~\ref{disp}).

As seen from Fig.~\ref{disp}, for all EM modes in the graphene ribbon the propagation length decays strongly close to the modal cutoff, being much smaller than the one in a 2D graphene sheet, and coincides with the latter for high frequencies. Between these two regions, each curve for $L/\lambda $ has a maximum at a finite frequency. To understand such a peculiar behavior let us first turn to a very interesting property of 2DGSPs. In the studied frequency range the real part of the conductivity (responsible for the dissipation) decreases as the frequency increases. However, due to increase of $\mathrm{Re}(q)$, the confinement increases so quickly with the frequency increase that GSPs become more absorptive, and consequently the propagation length of 2DGSPs decreases. Returning to the modes in the ribbons, in the region of high frequencies, where their $\mathrm{Re}(q)$ are large and, actually, coincide with $\mathrm{Re}(q^{2D})$, the confinement of the modes is high and their absorption increases with the frequency, exactly as in the case of 2DGSPs. For lower frequencies, where $\mathrm{Re}(q)$ is not so large, the confinement is not so sensible to the value of $q$ and the losses of the modes follow the character of the losses in a graphene sheet: they increase with the frequency decrease. Importantly, while $q$ decreases, there are also finite wavevector components in the in-plane directions across the ribbon that contribute to the confinement. Thus, the maxima in the propagation length corresponds to the optimum balance between strong plasmon confinement in a less lossy sheet (high frequencies) and smaller confinement in a more absorbing sheet (lower frequencies).

We now consider how the WGSP characteristics depend on different parameters. Their dependence on the relaxation energy is represented in Fig.~\ref{disp1} (a), showing that, while propagation length of the modes is very sensitive to  $E_{\tau}$ in this frequency regime, their wavevector is practically unaffected by it. This is related to the fact that the imaginary part of the conductivity is almost independent upon $\tau$, while the real part strongly depends on it, especially for lower frequencies. With the frequency increase, the GSPs losses become less effected by the relaxation time of electrons being more sensitive to the temperature.

With respect to the presence of a substrate, our calculations show that the dispersion relation of WGSP has the same structure as that shown in Fig.~\ref{disp} for the free standing case. Figure.~\ref{disp1}(b) renders, as a function of the  dielectric constant of the substrate ($\varepsilon_2$), the spectral value of the "asymptotes", i.e. the wavevector and propagation length for both the EGSP supported by a semi-infinite graphene sheet (discontinuous lines) and the 2DGSP (continuous lines). As $\varepsilon_2$ increases, the GSPs become more localized [$\mathrm{Re}(q)$ increases and, correspondingly, the propagation length decreases]. In the region of high frequencies, where the momentum is very large, $q\gg1$, the GSPs are less sensitive to the dielectric constant of the substrate.

To conclude, we have studied the dispersion characteristics of SP modes existing in graphene ribbons in the THz frequency range. We have shown that there are two types of SP modes: the waveguide-type, with the field concentrated along the whole area of the ribbon (in $x$-direction) and the edge modes, with the field concentrated on the rims of the ribbons, $x=\pm w/2$. The waveguide and edge modes are separated by a wavevector gap from each other. The number of SP modes supported by the ribbon increases as either the frequency or the ribbon width increase. The wavevector of the modes is not sensible to the relaxation time of charge carriers, but the propagation length is strongly affected. The SP modes can be tuned by changing the dielectric environment of the ribbon. The high localization of graphene EM edge modes can be useful for, for instance, bending of EM signals on subwavelength scales and enhancing the EM coupling between objects.

The authors acknowledge support from the Spanish MECD under Contract No. MAT2009-06609-C02 and Consolider Project ``Nanolight.es''. A.Y.N.
acknowledges the Juan de la Cierva Grant No. JCI-2008-3123.

\end{document}